%% file: article.tex
\documentclass[aps,
	prf,
	superscriptaddress,
	reprint,
	onecolumn,
	amsfonts,
	amssymb,
	amsmath,
]{revtex4-2}

\usepackage{graphicx}
\usepackage{color}

\input{macros.tex}

\begin{document}

\title{A mixing length model for arbitrary geometry: the case of parallel flows}

\author{Vincent Labarre}
\email[]{vincent.labarre@polytechnique.edu}
\affiliation{LadHyX (CNRS UMR 7646), Ecole Polytechnique, 91128 Palaiseau, France}

\author{Christophe Josserand}
\email[]{christophe.josserand@polytechnique.edu}
\affiliation{LadHyX (CNRS UMR 7646), Ecole Polytechnique, 91128 Palaiseau, France}

\author{Martine Le Berre}
\email[]{martinej.leberre@gmail.com}
\affiliation{ISMO (CNRS UMR 8214), Université de Paris-Saclay, 91405 Orsay, France}

\author{Romain Monchaux}
\email[]{romain.monchaux@ensta-paris.fr}
\affiliation{Institute of Mechanical Sciences and Industrial Applications, ENSTA-Paris, Institut Polytechnique de
	Paris, 828 Bd des Maréchaux, F-91120 Palaiseau, France}

\author{Luc Pastur}
\email[]{luc.pastur@ensta-paris.fr}
\affiliation{Institute of Mechanical Sciences and Industrial Applications, ENSTA-Paris, Institut Polytechnique de
	Paris, 828 Bd des Maréchaux, F-91120 Palaiseau, France}

\author{Yves Pomeau}
\email[]{yves.pomeau@gmail.com}
\affiliation{LadHyX (CNRS UMR 7646), Ecole Polytechnique, 91128 Palaiseau, France}

\begin{abstract}
	We present a phenomenological model for the mixing length used in turbulence models. It has the advantage of naturally accounting for the object's geometry while satisfying the standard symmetries of the Navier-Stokes equations. We employ the model to study channel flow and pipe flow. We calibrate the three model parameters to recover the damping in the viscous sub-layer, the log-law of the wall, and the outer region behaviors. Our model compares favorably to friction factor measurements in the pipe flow at high Reynolds numbers and gives analytical predictions of the mixing length for several canonical flows.	 
\end{abstract}

\maketitle

\section{Introduction}

Turbulence is fundamental in many natural and industrial flows such as in the atmosphere, for the wake behind solid bodies, or hydraulic systems. Despite the fact that the underlying fluid equations for turbulence, the Navier-Stokes equations,  fully describe turbulence. Unfortunately, these equations have no known analytical solutions except for particular cases, and their Direct Numerical Simulations (DNS) are often impossible due to the range of spatiotemporal scales involved.
One must model the effect of the small spatiotemporal scales to predict the flow. There exist numerous models, using different strategies \cite{pope2000turbulent}. In Reynolds Averaged Navier-Stokes (RANS) strategy aims to decompose the flows into a fluctuating part and a mean part and model the effect of the fluctuating part on the mean part. \\

We classify the RANS models by their complexity in modeling the Reynolds stress tensor $\bbsigma \equiv \mean{\uu' \otimes \uu'}$, $\uu'$ being the fluctuating velocity. Nowadays, Reynolds stress transport models, representing the evolution of the Reynolds stress tensor, are often used \cite{durbin2010ch7}. They have the advantage of representing nonlocal and anisotropic aspects of turbulence. Yet, they are complex and require fixing many free parameters on experimental or DNS data. One or two equations models like the Spalart-Allmaras \cite{spalart1992one}, the $k-\varepsilon$ \cite{hanjalic1972reynolds}, $k-\omega$ \cite{wilcox2008formulation}, and the Shear Stress Transport model \cite{menter1994two}, are still widely used in engineering. They represent nonlocal aspects of turbulence by modeling the transport of characteristic scalar(s) of turbulence, like the kinetic energy $k$ and dissipation $\varepsilon$. A significant challenge in employing complex models (or DNS) lies in accurately resolving the viscous sublayer near object boundaries. The height of this sublayer diminishes as the Reynolds number increases, augmenting the computational cost. \\

The simplest RANS models are based on the mixing length hypothesis. By analogy to molecular viscosity, the Reynolds stress tensor is linked to the local mean velocity gradient by $\bbsigma = -\ellm^2 \left| \bar{\bar{S}} \right| \bar{\bar{S}}$ (or a similar expression) where $\ellm$ is the mixing length, that depends on the position, and $\bar{\bar{S}}$ is the mean velocity shear \cite{prandtl1925bericht}. This description assumes that momentum transport is mediated by fluid parcels of mean free path $\ellm$. Yet, unlike viscous diffusion, there is no obvious intrinsic length scale in turbulence, making this hypothesis questionable. Because there are no fluctuations at boundaries, the mixing length is zero with an exponential decay \cite{vandriest1956}. Close to boundaries, the mixing length is proportional to the distance to the boundary \cite{karman1930mechanische}. Far from boundaries, one assumes the mixing length to be proportional to the dimension of the flow where the shear producing turbulence is sustained. At the same time, the mixing length is expected to be smaller than the flow size because large eddies transfer energy to smaller eddies \cite{richardson2007weather, kolmogorov1991local, frisch1995turbulence}. \\

Mixing length models are much easier to use because of their low computational and calibration costs. They are still used to understand fundamental aspects of turbulence in several applications including wall-bounded flows \cite{subrahmanyam2022universal}, or far wake behind bodies \cite{hutchinson2021prandtl}. Because of their low computational cost, they also allow refinement of the numerical grid near objects' boundaries compared to more elaborated models. This approach also appeals to parametric studies, where complex models have prohibitive computational costs. The mixing length is more easily defined for flows with simple geometries or when turbulent quantities are known (e.g. $\ellm \propto k^{3/2}/\varepsilon$). By opposition, the mixing length is not easily defined for complex geometries or when $k$ and $\varepsilon$ are unknown. Even for parallel flows, the mixing length is not trivial, and several models have been proposed. A sophisticated ansatz for the mixing length allows the fitting of the velocity profiles for pipe flow experiments over three orders of magnitude in the Reynolds number \cite{cantwell2019universal}. For the channel flow, a symmetric version of the mixing length has been proposed \cite{sun2023turbulent}. In \cite{she2017quantifying,chen2018quantifying}, the authors employ a Lie-group analysis of the RANS equations to determine the mixing length. The method yields excellent results but is limited to parallel flows. \\

This study proposes an approach for determining the mixing length for turbulent flows across all geometries, offering a phenomenological yet physically grounded solution that obviates the need for solving additional equations for turbulent quantities. The model allows us to recover the van Driest damping \cite{vandriest1956}, the ``Log-law of the wall'' \cite{karman1930mechanische}, and the saturation of the mixing length far from boundaries. \\

The remaining of the manuscript is as follows. In section~\ref{sec:Model}, we present our mixing length model. We present our results in section~\ref{sec:Results}. 
In the first subsection~\ref{subsec:Channel}, we compare our model to DNS for channel flow. In a second subsection~\ref{subsec:Pipe}, we compare our model to friction coefficient measurements in pipe flows. We give concluding remarks in section~\ref{sec:Conclusions}.

\section{The model \label{sec:Model}}

Our starting point is the Navier-Stokes equations for a Newtonian fluid in an inertial frame, assuming incompressibility of the flow, neglecting thermal effects and viscosity variation. Then, the dynamical equations write
\begin{align}
	\label{eq:Incompressibility}
	\bnabla \cdot \vv &= 0, \\
	\label{eq:NavierStokes}
	\p_t \vv + \vv \cdot \bnabla \vv &= - \frac{1}{\rho} \bnabla p + \nu \Delta \vv
\end{align}
where $\xx$ is the position vector, $\vv(\xx,t)$ is the velocity, $\rho$ is the constant density, $p(\xx,t)$ is the pressure (plus eventually the potential of conservative forces), and $\nu$ is the kinematic viscosity. The last equations give the velocity field in the fluid domain $\V$. On the boundaries $\p \V$ of the domain, i.e. at the contact of objects, the flow satisfies no-slip boundary conditions: the velocity of the fluid is equal to the object's velocity. We consider a flow of typical velocity $U$ and typical size $L$. The Reynolds number is $Re = UL/\nu$ and is the only relevant control parameter of the flow apart from the geometry. We use the rescaling $\vv \rightarrow U \vv$, $p \rightarrow \rho U^2 p$, and $\xx \rightarrow L \xx$. When $Re \gg 1$, the flow starts to be chaotic, then turbulent, and a statistical description of the velocity field is foreseen. In this study, we employ the Reynolds decomposition of the velocity, $\vv(\xx,t) = \uu(\xx,t) + \uu'(\xx,t)$, where $\uu(\xx,t) = \mean{\vv(\xx,t)}$ represents an ensemble average (which is equal to a temporal average if the ergodic assumption is satisfied), and $\uu'$ is the fluctuating part. The same decomposition applies to the pressure field, namely $p(\xx,t) = \mean{p(\xx,t)} + p'(\xx,t)$. Since we will not discuss the pressure fluctuations, we will note $\mean{p(\xx,t)} \rightarrow p(\xx,t)$ in the following to keep compact notations. Then, the equations describing the evolution of the mean velocity are
\begin{align}
	\label{eq:IncompressibilityMean}
	\bnabla \cdot \uu &= 0, \\
	\label{eq:NavierStokesMean}
	\p_t \uu + \uu \cdot \bnabla \uu &= - \bnabla p + \frac{1}{Re} \Delta \uu - \bnabla \cdot \bar{\bar{\sigma}}
\end{align}
where 
\begin{equation}
	\bbsigma = \mean{\uu' \otimes \uu'}
\end{equation}
is the Reynolds Stress Tensor (RST). It is clear from the last equations that we need a model for the RST to obtain a closed system of equations for the mean velocity $\uu(\xx,t)$. It is the so-called closure problem. The velocity fluctuations are zero at the boundaries, implying that all components of $\bar{\bar{\sigma}}(\xx,t)$ and all their spatial derivatives are zero at the boundaries $\p \V$, i.e. 
\begin{equation}
	\label{eq:BCRST}
	\bbsigma(\xx) = 0 ~~~~ \text{and} ~~~~ \nn \cdot \bnabla \bbsigma(\xx) = \bzero ~~~~ \forall \xx \in \p \V,
\end{equation}
where $\nn$ is the unitary vector orthogonal to the boundaries. The last conditions are strong constraints on the RST. We consider a mixing length model, that reads
\begin{equation}
	\label{eq:RST}
	\bbsigma = - \ellm^2 \left| \bnabla \uu + (\bnabla \uu)^{\rm T} \right| \left( \bnabla \uu + (\bnabla \uu)^{\rm T} \right)
\end{equation}
where $\ellm(\xx)$ is the dimensionless mixing length (scaled by $L$). \\

Inspired by recent studies \cite{josserand2020scaling,pomeau2021turbulence,pomeau2021turbulent}, which propose simple nonlocal RST models, we try to construct a geometry dependant mixing length that takes into account nonlocal aspects of turbulence. The model has to satisfy the standard symmetries: time reversal and translation, space reversal and translational invariances, rotational invariances, and Galilean invariance. Due to the boundary conditions (\ref{eq:BCRST}) and the fact that the velocity gradient is generally not zero at boundaries, the mixing length has to vanish at boundaries. We will also assume that the tangential derivative of the mixing length has to vanish at boundaries so
\begin{equation}
	\label{eq:BCML}
	\ellm(\xx) = 0 ~~~~ \text{and} ~~~~ \nn \cdot \bnabla \ellm(\xx) = 0 ~~~~ \forall \xx \in \p \V,
\end{equation}
To model $\ellm(\xx)$, we use the following partial differential equation:

\begin{equation}
	\label{eq:Model}
	- \alpha^2 \Delta^2 \ellm  + \beta^2 \Delta \ellm - \gamma^2 \ellm + 1 = 0,
\end{equation}
where $\alpha$, $\beta$ and $\gamma$ are dimensionless constants that will we calibrate in subsection \ref{subsec:Channel}. Despite being an empirical equation, we can give some physical meanings to all the terms:
\begin{itemize}
	\item $-\alpha^2 \Delta^2 \ellm$ is important at small length scales, and can be used to impose the two boundary conditions (\ref{eq:BCML}). As we will see, $\alpha$ depends on $Re$ (i.e. the viscosity). To anticipate the value of $\alpha$, we can note that, in turbulent flows, the viscosity is important mostly in the viscous sub-layers near the boundaries, whose size scales as $1/Re$. For this reason, we expect that $\alpha \propto 1/Re$ for sufficiently large $Re$. 
	\item $\beta^2 \Delta \ellm$ is important at intermediate length scales. If $\alpha=0$, it allows to impose $\ellm(\xx) = 0$ or $\nn \cdot \bnabla \ellm(\xx) = 0$, but not the two conditions simultaneously. Therefore, it makes the flow dependent on the geometry, but we expect $\beta$ to be independent of viscosity.  
	\item $-\gamma^2 \ellm$ is important at large length scales. If $\alpha=\beta=0$, it leads to a constant $\ellm$ and imposes no boundary conditions. Therefore, this term would represent the flow in regions where the precise shape of the object and the viscosity are unimportant. \\
\end{itemize} 
The alternate signs and the physical relevance of equation (\ref{eq:Model}) will become clearer when discussing the links with existing models for the channel flow in subsection \ref{subsec:Channel}. We anticipate that $\alpha \ll 1$ and $\beta, \gamma \sim 1$ for turbulent flows. Note that (\ref{eq:Model}) satisfies all the standard symmetries because of the use of the Laplacian operator, but these symmetries can be broken by boundary conditions (\ref{eq:BCML}). Using the Laplacian operator is also natural when we want to account for the non-locality of turbulence. \\

Using equations involving the Laplacian operator to model turbulence reminds elliptic relaxation models which use an elliptic equation to model the redistribution term in the RST transport equation \cite{durbin1991near,durbin1993reynolds,pope2000turbulent}. Yet, our model is different since we use an equation to model the mixing length without using turbulent quantity fields. It has the advantage of recovering wall damping while keeping our model simple and not using too many adjustable parameters and additional equations. However, models that resolve the transport of turbulent quantities, like the $k-\varepsilon$ model \cite{hanjalic1972reynolds}, take into account the effect of flow on the mixing length, while our model does not account for this effect. More precisely, the (global) Reynolds number and the boundary conditions are the only characteristics that will modify the mixing length in our model.

\section{Results \label{sec:Results}}

\begin{figure}
	\includegraphics[scale=1]{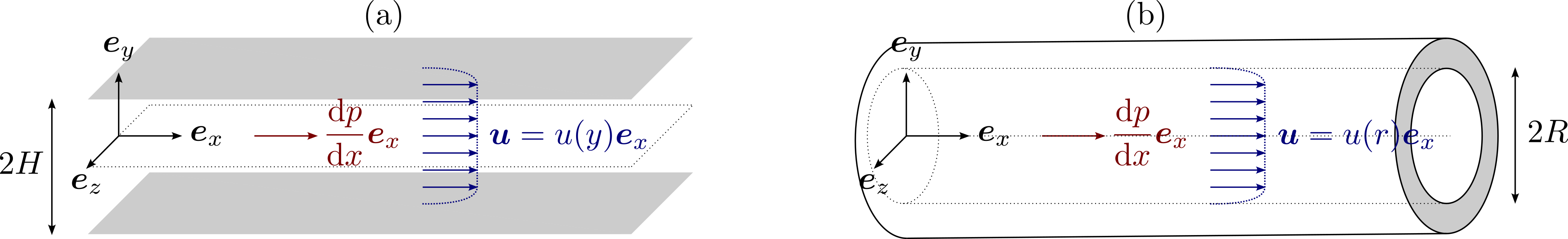}
	\caption{Sketches of (a) channel flow described in subsection \ref{subsec:Channel} and (b) pipe flow described in subsection \ref{subsec:Pipe}. \label{fig:Sketchs}}
\end{figure}

\subsection{Channel flow \label{subsec:Channel}}

The geometry under consideration in this subsection is a statistically steady, turbulent channel flow in between two parallel (and infinite) plates located, respectively, at the elevation $y=-1$ and $y=1$, see Fig.\ref{fig:Sketchs}(a). The flow is driven by a constant uniform pressure gradient $- (\diff p / \diff x) ~ \eex \equiv \eex$. Because of the symmetries, if time or ensemble averaged, we can think of statistics that depend on $y$ only. The mean velocity is $\uu = u(y) \eex$. The only component of the RST we shall deal with is $\sigma_{xy}(y) = \mean{u_x' u_y'}$. Note that choosing $|(\diff p / \diff x)| = 1$ in dimensionless units is equivalent to rescaling the Navier-Stokes equations using the typical velocity $U = \sqrt{-(\diff p/\diff x) H / \rho}$, $H$ being the distance between the center and the plates in physical units. The averaged stress balance along $\eex$ yields the equation  
\begin{equation}
	\label{eq:MomentumChannel}
	\der{\sigma_{xy}}{y} - \frac{1}{Re} \dfrac{\diff^2 u}{\diff y^2} = 1,
\end{equation}
and the boundary conditions read $u(\pm 1)=0$, $\sigma_{xy}(\pm 1)=0$, and $\der{\sigma_{xy}}{y}(\pm 1)=0$. From equations (\ref{eq:RST}-\ref{eq:Model}), we obtain the mixing length and the RST:
\begin{align}
	\label{eq:MLChannel}
	\ellm(y) &= \frac{1}{\gamma^2} \frac{r_+ \left( e^{-2r_+} - 1 \right) \left( e^{r_- (y-1)} + e^{r_- (-y-1)} \right) + r_- \left( 1 - e^{-2r_-} \right) \left( e^{r_+ (y-1)} + e^{r_+ (-y-1)} \right) }{r_+ \left( 1-e^{-2r_+} \right) \left( 1+e^{-2r_-} \right) + r_- \left( 1+e^{-2r_+} \right) \left( e^{-2r_-}  - 1 \right) } + \frac{1}{\gamma^2}, \\
	\label{eq:RSTChannel}
	\sigma_{xy} &= -\ellm^2 \left| \der{u}{y} \right| \der{u}{y},
\end{align} 
where 
\begin{equation}
	\label{eq:Roots}
	r_\pm^2 = \frac{\beta^2 \pm \beta^2 \sqrt{1 - 4 \alpha^2 \gamma^2 /\beta^4}}{2\alpha^2} \geq 0.
\end{equation}
For small $\alpha$, the two roots (\ref{eq:Roots}) are positive because of the alternate of signs in equation (\ref{eq:Model}). It corresponds to monotonous damping/growth of the mixing length. For example, if we have used $-\alpha^2 \rightarrow \alpha^2$ in equation (\ref{eq:Model}), $r_-^2$ (\ref{eq:Roots}) would be negative, corresponding to unphysical oscillating and/or negative mixing length. Assuming that $\alpha \rightarrow 0$ as $Re \rightarrow \infty$, we obtain
\begin{equation}
	\label{eq:RootsAsymp}
	r_+(Re \rightarrow \infty) \rightarrow r_{+,\infty} \equiv \frac{\beta}{\alpha} \gg 1 ~~~~ \text{and} ~~~~ r_- (Re \rightarrow \infty) \rightarrow r_{-,\infty} \equiv \frac{\gamma}{\beta} \sim 1.
\end{equation}
To better understand the physical meaning of the constants $\alpha, \beta$ and $\gamma$ or, equivalently, $r_\pm$ and $\gamma$, we can consider the following asymptotics, corresponding to three different flow regions. \\

\paragraph{Viscous sub-layer: $1 \pm y \rightarrow 0$:}  
{
	Very close to the boundary, most of the variation of $\ellm(y)$ (\ref{eq:MLChannel}) are due to the fast exponential damping at rate $r_+$. We link this behavior to the so-called van Driest damping \cite{vandriest1956}, which consists of adding an exponential damping coefficient $(1 - e^{-Re (1 \pm y) / A})$ to the mixing length, $A$ being a damping coefficient. This inspired guess allows us to find a better agreement near the walls. In our model, we can recover the same behavior for large $Re$ near the wall by fixing  
	\begin{equation}
		\label{eq:Constraint1}
		r_{+,\infty} = \frac{\beta}{\alpha} = \frac{Re}{A}.
	\end{equation}
	It is the first constraint on our coefficients. \\
}

\paragraph{Log-region: $(1 \pm y)r_+ \gg 1$ and $(1 \pm y)r_- \ll 1$:}
{
	Far from the boundary (in viscous scale) but far from the center, we obtain the following asymptotic for the mixing length (\ref{eq:MLChannel})
	\begin{equation}
		\ellm(y) \simeq \frac{r_- \left( e^{2r_-} - 1\right)}{\gamma^2 \left( e^{2r_-} + 1\right)} \left( 1 \pm y \right).
	\end{equation}	
	We link this behavior to the von Kármán model, that is $\ellm(y) = \kappa (1 \pm y)$, $\kappa \simeq 0.41$ being the von Kármán constant. Using the large $Re$ limit (\ref{eq:RootsAsymp}), it gives the second constraint on our coefficients
	\begin{equation}
		\label{eq:Constraint2}
		\frac{\gamma}{\beta} \frac{1-e^{-2\gamma/\beta}}{1+e^{-2\gamma/\beta}} = \kappa \gamma^2.
	\end{equation} \\
}

\paragraph{Outer region: $y \ll 1$:}
{
	To obtain the third constraint on our coefficients, we fix 
	\begin{equation}
		\label{eq:Constraint3}
		\frac{1}{\gamma^2} \equiv C
	\end{equation}
	where $C$ is a numerical constant related to the value of the mixing length in the middle plane $y=0$. For the sphere case, it can be shown that $C$ corresponds to the value of the mixing length at infinity. Its typical value is $\simeq 0.14-0.18$ for several flows \cite{greenshields2022notes}. \\
}

For a given set of parameters $(A,\kappa,C)$ and a given Reynolds number, we compute $(\alpha,\beta,\gamma)$ using (\ref{eq:Constraint1},\ref{eq:Constraint2},\ref{eq:Constraint3}). We solve (\ref{eq:Constraint2}) numerically using the Newton method. Note that only $\alpha$ depends on $Re$, while $(\beta,\gamma)$ depend only on $(\kappa,C)$. Our model is therefore well constrained by robust observations in turbulent flows. Then, our model reproduces both the van Driest damping near the wall and the log-law of the wall, as we will check numerically. It also has the advantage of being formulated for arbitrary geometry. Another feature of the model is the presence of a critical number, under which the roots (\ref{eq:Roots}) become imaginary numbers. The threshold is given by the condition $1 - 4 \alpha^2 \gamma^2 / \beta^4 = 0$ which, using (\ref{eq:Constraint1}), gives
\begin{equation}
	\label{eq:Rec}
	Re =  \frac{2A \gamma}{\beta} = \Rec(A,\kappa,C)
\end{equation}	
where $\Rec$ is a critical Reynolds number that we compute numerically. For $Re<\Rec$, the roots (\ref{eq:Roots}) are imaginary, so unphysical, while $Re>\Rec$, the two roots are real. We interpret $\Rec$ as the threshold for the sub-critical transition to turbulence, first observe by Reynolds \cite{reynolds1883experimental}. Note that $\Rec$ depends on $(A,\kappa,C)$, giving another physical constraint on our model. We will show that our prediction for $\Rec$ is consistent with pipe flow experiments in subsection \ref{subsec:Pipe}.  \\

After integration of (\ref{eq:MomentumChannel}), using (\ref{eq:MLChannel}-\ref{eq:RSTChannel}), and considering that $\diff u/\diff y \leq 0$ for $y \in [0,1]$, we can rewrite
\begin{equation}
	\label{eq:ImpulsionChannel}
	- \ellm^2 \left( \der{u}{y} \right)^2 + \frac{1}{Re} \der{u}{y} + y = 0 ~~~~ \forall y \in [0:1].
\end{equation}
It is easy to derive the equation for the velocity gradient
\begin{equation}
	\label{eq:FinalChannel}
	\der{u}{y} = \frac{1 - \sqrt{1 + 4 Re^2 \ellm^2 y}}{2Re \ellm^2} ~~~~ \forall y \in [0:1].
\end{equation}
We reconstruct the velocity field by numerical integration using its derivative, the boundary condition $u(1)=0$, and the trapezoidal integration rule. We employ $N/2$ linearly spaced points for $y \in [0;0.95]$ and $N/2$ logarithmic spaced points for $y \in [0.95;1]$ (with more points near $y=1$). It allows us to resolve accurately the viscous sub-layer and the outer layer while keeping a reasonable number of points in the logarithmic region. \\

\begin{figure}
	\includegraphics[width=\linewidth]{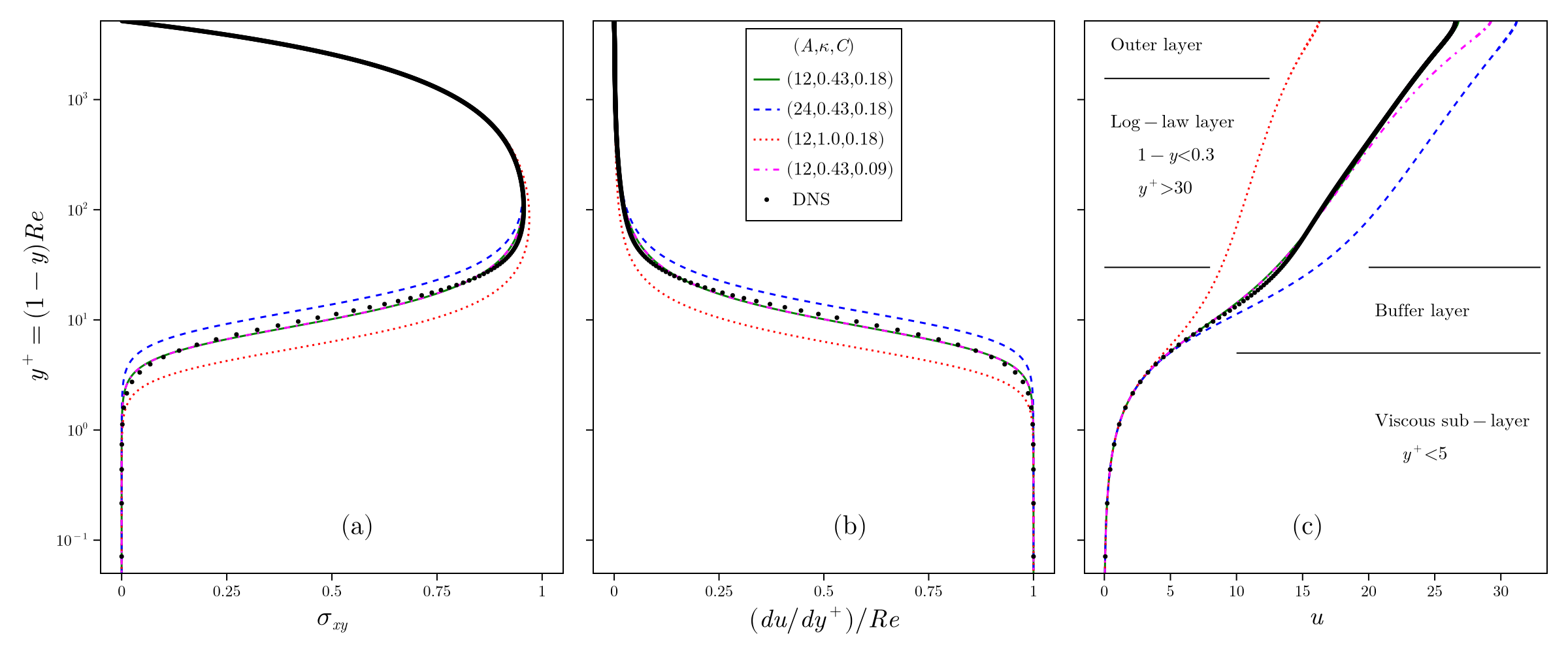}
	\caption{Comparison between the model and the DNS from John Hopkins Turbulence Database \cite{lee2015direct}. (a) Parietal component of the RST, (b) derivative of the velocity, and (c) velocity profiles. $y^+$ is the wall distance in viscous length-scale. The names and definitions of the different regions are given in panel (c), see \cite{pope2000turbulent}. Each curve corresponds to a given value of the parameters $(A,\kappa,C)$. \label{fig:ChannelvsDNS}}
\end{figure}

In Fig.\ref{fig:ChannelvsDNS}, we compare our model to channel flow DNS of the John Hopkins Turbulence Database for $Re \simeq 5200$ \cite{lee2015direct}. For $(A,\kappa,C)=(12,0.43,0.18)$ (full green curve), we see a very good agreement with DNS, except for small deviations in the buffer layer. These deviations are explained by the fact that our model predicts, as for the van Driest model, $\sigma_{xy} \propto y^{+4}$ instead of the correct behavior $\sigma_{xy} \propto y^{+3}$ near the wall \cite{pope2000turbulent}. Note that $A$ and $\kappa$ have slightly different values than in the literature. When we increase $A$ (dashed blue curve), the viscous sub-layer thickness increases, as expected. If we increase $\kappa$ to unity (dotted red curve), the slope of the log-law layer decreases, as shown in panel (c). When we decrease $C$ (dashed-dotted magenta line), the profiles remain similar to the outer layer, where differences are visible in the velocity profile. In the outer region, the mixing length is $\ellm(y) = \ellm(0) + O(y^2) \propto C$, so we can integrate (\ref{eq:FinalChannel}) by considering a constant mixing length. The result is
\begin{equation}
	\label{eq:Wake}
	u(y) \simeq u(0) + \frac{y}{2Re\ellm^2(0)} - \frac{2}{3\ellm(0)} \left[ \left( \frac{1}{4Re^2\ellm^2(0)} + y \right)^{3/2} - \left( \frac{1}{4Re^2\ellm^2(0)} \right)^{3/2} \right]
\end{equation}
Therefore, the parameters $A$, $\kappa$, and $C$ have clear effects on the viscous, log-law, and outer layers. \\

\begin{figure}
	\includegraphics[width=\linewidth]{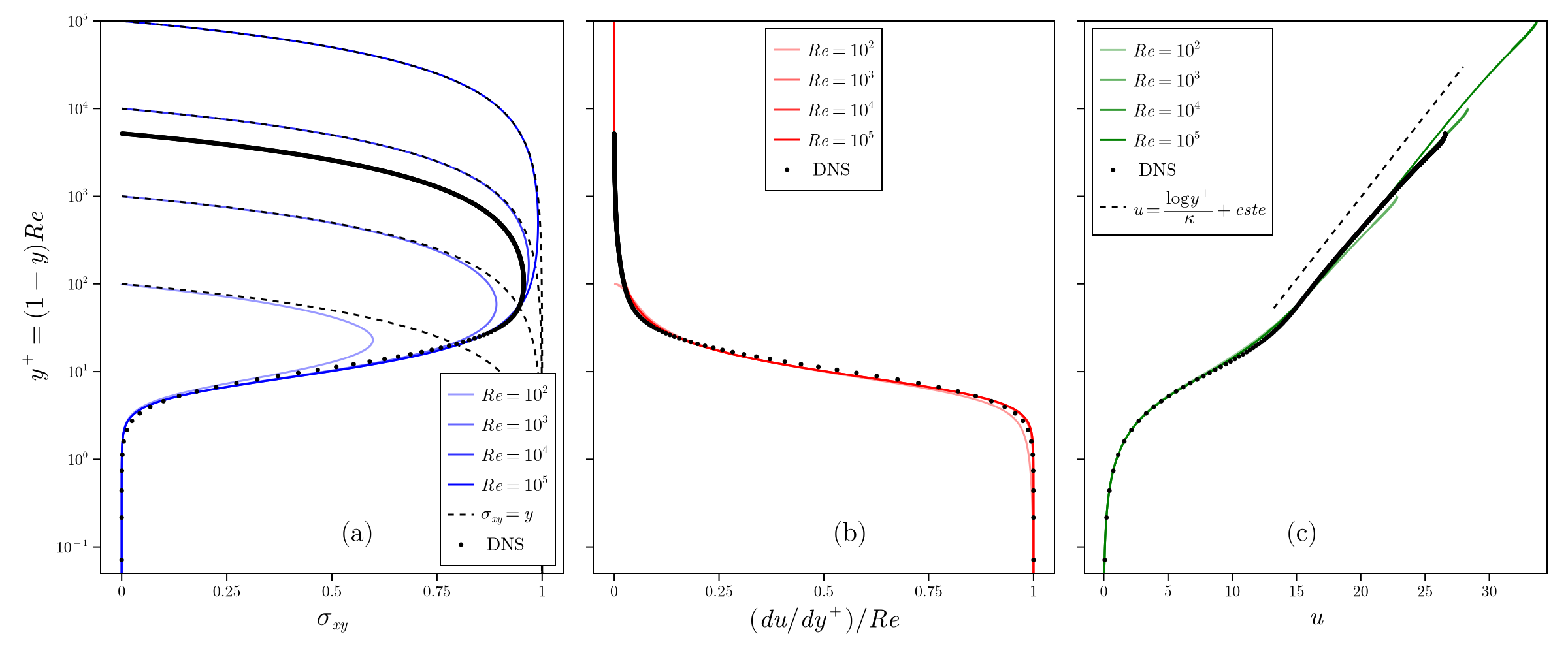}
	\caption{(a) Parietal component of the RST, (b) derivative of the velocity, and (c) velocity profiles for different $Re$ and $(A,\kappa,C) = (0.12, 0.43, 0.18)$. In panel (a), the dashed lines correspond to the fully turbulent prediction $\sigma_{xy}(y) = y$ for the different $Re$. In panel (c), the dashed line corresponds to the log-law with $\kappa=0.43$. $y^+$ is the wall distance in viscous length-scale. We show the profiles from a DNS with $Re \simeq 5200$ \cite{lee2015direct} for comparison. \label{fig:ChannelvsRe}}
\end{figure}

In Fig.\ref{fig:ChannelvsRe}, we show the profiles $\sigma_{xy}$, $(du/dy^+)/Re$, and $u$ for various Reynolds numbers. As we increase $Re$, we observe a development of the inertial region, characterized by a longer range for $y^+$. In panel (a), we see that $\sigma_{xy}(y)$ gets closer to the turbulent prediction $\sigma_{xy}(y) = y$ when increasing $Re$, for $y^+$ sufficiently large. In panel (b), we observe that the derivatives of the velocity profiles are very similar near the wall for $Re \geq 10^3$. Only the simulation with $Re=10^2$ is significantly affected by the viscosity in all the domain. In panel (c), the velocity profiles show a region compatible with the log-law, as anticipated. This region, characterized by $\ellm(y) \simeq \kappa (1 \pm y)$, gets larger as we increase $Re$. For the largest $y^+$, the velocity profiles deviate from the log-law because $\ellm$ is no longer linear and saturated to a constant value. \\

\subsection{Pipe flow \label{subsec:Pipe}}

\begin{figure}
	\includegraphics[width=\linewidth]{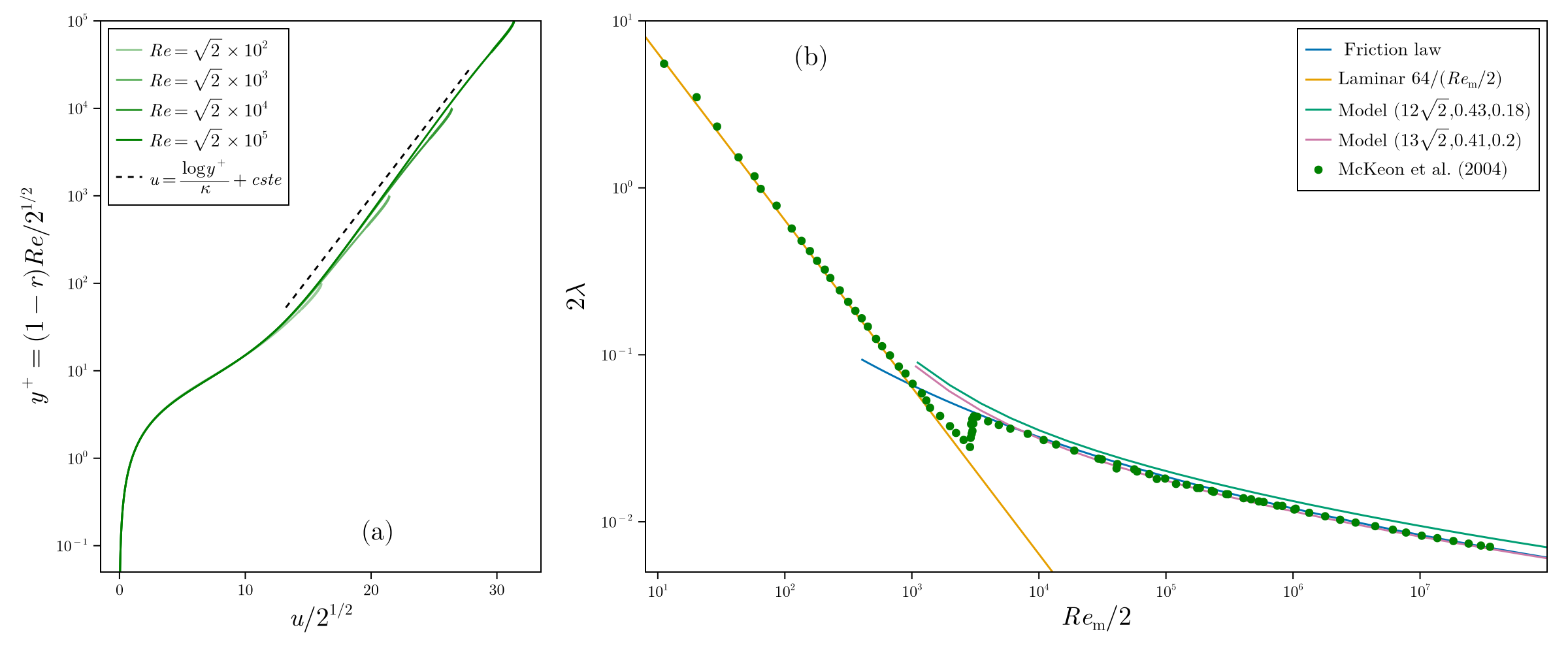}
	\caption{(a) Velocity profile for the pipe flow and different $Re$. The parameters of the model are $(A,\kappa,C)=(12\sqrt{2},0.43,0.18)$. (b) Friction coefficient $\lambda$ (\ref{eq:Friction}) as a function of the Reynolds number based on the average velocity $\Rem$ (\ref{eq:Rem}). We show the laminar prediction, Prandtl's friction law (\ref{eq:PrandtlFriction}), the predictions for our model for two sets of parameters $(A,\kappa,C)$, and experimental data \cite{mckeon2004friction}. \label{fig:Pipe}}
\end{figure}

We now consider a flow in an infinite pipe of axis $\eex$ and circular section of radius unity, as shown in Fig.\ref{fig:Sketchs}(b). It is driven by a constant uniform pressure gradient along $x$, $-\diff p / \diff x = 2$. In this case, the various quantities involved depend on the radial distance only, $r = \sqrt{y^2 + z^2}$ and we shall consider the component $\sigma_{xr}(r)$ of the RST. Note that choosing $|(\diff p / \diff x)| = 2$ in dimensionless units is equivalent to rescaling the Navier-Stokes equations using the typical velocity $U = \sqrt{- (\diff p / \diff x) R / (2 \rho)}$, $R$ being the radius of the pipe in physical units. The average momentum equation along the pipe axis is
\begin{equation}
	\label{eq:MomentumPipe}
	\der{(r \sigma_{xr})}{r} - \frac{1}{Re} \dfrac{\diff}{\diff r} \left( r \der{u}{r} \right) = 2 r,
\end{equation}
and the boundary conditions read $u(1)=0$, $\sigma_{xr}(1)=0$ and $\der{\sigma_{xr}}{r}(1)=0$. From equations (\ref{eq:RST}-\ref{eq:Model}), we obtain the mixing length and the RST:
\begin{align}
	\label{eq:MLPipe}
	\ellm(r) &= \frac{1}{\gamma^2} \frac{r_- I_1(r_-) I_0(r_+ r) - r_+ I_1(r_+) I_0(r_- r)}{r_+ I_0(r_-) I_1(r_+) - r_- I_0(r_+) I_1(r_-)} + \frac{1}{\gamma^2}, \\
	\label{eq:RSTPipe}
	\sigma_{xr} &= -\ellm^2 \left| \der{u}{r} \right| \der{u}{r},
\end{align} 
where $I_n(r)$, $n \in \mathbb{N}$, are the modified Bessel functions of the first kind, and the roots $r_\pm$ are still given by equation (\ref{eq:Roots}). $I_0(r)$ and $I_1(r)$ are difficult to evaluate numerically for large $r$. For this reason, we use the asymptotic $I_0(r) \simeq I_1(r) \simeq e^r/\sqrt{2\pi r}$ \cite{watson1995treatise} to derive the following expression for large $r_+$: 
\begin{equation}
	\ellm(r) \simeq \frac{1}{\gamma^2} \frac{r_- I_1(r_-) e^{-r_+(1-r)}/\sqrt{r} - r_+ I_0(r_- r)}{r_+ I_0(r_-) - r_- I_1(r_-)} + \frac{1}{\gamma^2}. \\
\end{equation} 
We use the previous expression instead of (\ref{eq:MLPipe}) for $r_+>500$. After integration of (\ref{eq:MomentumPipe}), using (\ref{eq:MLPipe}-\ref{eq:RSTPipe}), and considering that $\diff u/\diff r \leq 0$ for $r \in [0,1]$, we rewrite
\begin{equation}
	\label{eq:FinalPipe}
	- \ellm^2 \left( \der{u}{r} \right)^2 + \frac{1}{Re} \der{u}{r} + r = 0 ~~~~ \forall r \in [0:1],
\end{equation}
with $u(1)=0$. This equation is the same as for the channel flow (\ref{eq:ImpulsionChannel}) after replacing the coordinate $y$ by $r$. The velocity gradient satisfies the equation (\ref{eq:FinalChannel}) except that the mixing length is different. We can therefore exploit the same numerical method to solve equation (\ref{eq:FinalPipe}). Yet, to compare to the channel flow and other studies, we need to rescale the velocity and $Re$ by a factor $1/\sqrt{2}$, and $\alpha \propto 1/Re$ by a factor $\sqrt{2}$. In Fig.\ref{fig:Pipe}(a), we show the velocity profile in the pipe flow for different $Re$. It is similar to the channel flow shown in Fig.\ref{fig:ChannelvsRe}(c). In particular, we observe a region compatible with the log-law for sufficiently large $Re$. \\

To characterise friction in pipe, one often introduces the friction factor, which is defined as \cite{pope2000turbulent}
\begin{equation}
	\lambda = \left. \left( - \der{p}{x} 2R \right) \right/ \left( \frac{1}{2} \rho \Um^2 \right),
\end{equation} 
where $\Um$ is the average velocity. In dimensionless units, it reads
\begin{equation}
	\label{eq:Friction}
	\lambda = \frac{8}{\Um^2},
\end{equation}
where $\Um$ is now the average velocity scaled by $U = \sqrt{- (\diff p / \diff x) R / (2\rho)}$. Still in dimensionless unit, the Reynolds number based on the pipe diameter $2R$ and the mean velocity $\Um$ is 
\begin{equation}
	\label{eq:Rem}
	\Rem = 2 Re \Um ~~~~ \text{with} ~~~~ \Um = 2 \int\limits_0^1 ~ u(r) ~ r ~ \diff r.
\end{equation}
For a laminar flow, the mean velocity is $\Um = \Um^{({\rm L})} = Re/4$ such that $\Rem = \Rem^{({\rm L})} = Re^2/2$ and $\lambda = \lambda^{({\rm L})} = 64/\Rem$. \\

We plot the friction coefficient as a function of $\Rem$ in Fig.\ref{fig:Pipe}(b). Because of our choice for the typical velocity, we need to multiply $\lambda$ by a factor $2$ and divide $\Rem$ by two to compare to the experimental data of \cite{mckeon2004friction}. For small $\Rem$, the experimental points follow the laminar prediction $\lambda = 64 / \Rem$. For large $\Rem$, the friction factor is well fitted by the friction law for smooth pipes \cite{mckeon2004friction}, obtained by solving the implicit equation \cite{pope2000turbulent}
\begin{equation}
	\label{eq:PrandtlFriction}
	\frac{2}{\sqrt{\lambda}} - \frac{\log \left[ \Rem \sqrt{\lambda/2}  \right]}{\kappa} + \frac{\log 32}{2\kappa} - B = 0
\end{equation}
where $\kappa=0.41$ and $B=5.1$ is the offset coefficient in the log-law (i.e. $u \simeq \frac{1}{\kappa} \log y^+ + B$ in the log-region). Formulae similar to (\ref{eq:PrandtlFriction}) were also introduced in other studies \cite{colebrook1939turbulent, moody1944friction}. We show the predictions of our model for two sets of parameters $(A,\kappa,C)$ and $Re \geq \Rec$ (\ref{eq:Rec}). For these values of the parameters, the critical Reynolds number (\ref{eq:Rec}) is $\Rec \simeq 80$. Considering a laminar velocity profile, it corresponds to a transition at $\Rem = \Rec^2/2 \simeq 3.2 \times 10^3$. Therefore, our model predicts a transition around $\Rem/2 \simeq 1.6 \times 10^3$, slightly before the one observed in the experimental data \cite{mckeon2004friction}. For $(A,\kappa,C) = (12\sqrt{2},0.43,0.18)$, which corresponds to the values chosen for the channel flow in subsection \ref{subsec:Channel}, the model slightly overestimate the value of the friction factor. Changing the parameter to $(A,\kappa,C) = (13\sqrt{2},0.41,0.2)$ allows us to find an agreement comparable to the friction law.

\section{Conclusions \label{sec:Conclusions}}

We have introduced a phenomenological model for the mixing length. The model satisfies all the standard symmetries of the Navier-Stokes equations, except when boundary conditions break it due to the presence of objects. We have compared the predictions of our model for two parallel wall flows. Our model has 3 adjustable parameters, allowing us to fit well the velocity profile of the channel flow by adapting the behavior in the viscous sub-layer, the Log-region, and the outer region. Our model predicts a laminar-turbulent transition at a critical Reynolds number and a friction factor consistent with the pipe flow experiments, with only small adjustment of the parameters when compared to the channel flow. \\

Even though our model gives satisfactory results for parallel flows, it is worth mentioning that it does not account for the effect of the flow field structure on the mixing length. Instead, the mixing length depends only on the global Reynolds number and the geometry through a linear partial differential equation, which is disputable. Yet, the model may be interesting when more complex models like Reynolds stress transport models and large eddy simulations are too expansive, e.g. for resolving viscous-sub-layer, dealing with multiple objects, studying the far wake, and performing parametric studies over a large Reynolds number range. It also has a conceptual interest, since it attempts to give a universal definition of the mixing length, i.e. for arbitrary geometry. \\

We plan to check our model for open flows (wakes and jets). It is simple enough to allow analytical treatments in simple geometries. Particularly, we obtain an analytical solution for the mixing length in the case of a flow around a sphere.

\begin{acknowledgments}
	This work was partially supported by Agence de l'Innovation de D\'efense (AID) - via Centre Interdisciplinaire d'Etudes pour la D\'efense et la S\'ecurit\'e (CIEDS) - (project 2022 - SILTURB). We thank Bérengère Dubrulle, Sergio Rica, Christos Vassilicos, and Rémi Zamansky for fruitful discussions.
\end{acknowledgments}

\bibliography{biblio}

\end{document}

%% file: macros.tex
\newcommand{\V}{\mathcal{V}}

\newcommand{\eex}{\boldsymbol{e}_x}

\newcommand{\xx}{\boldsymbol{x}}

\newcommand{\nn}{\boldsymbol{n}}
\newcommand{\vv}{\boldsymbol{v}}
\newcommand{\uu}{\boldsymbol{u}}

\newcommand{\bnabla}{\boldsymbol{\nabla}}
\newcommand{\bzero}{\boldsymbol{0}}

\newcommand{\ellm}{\ell_{\rm m}}

\newcommand{\bbsigma}{\bar{\bar{\sigma}}}

\newcommand{\p}{\partial}
\newcommand{\diff}{\mathrm{d}}
\newcommand{\der}[2]{\dfrac{\diff #1}{\diff #2} }

\newcommand{\Rec}{Re_{\rm c}}

\newcommand{\mean}[1]{ \left\langle #1 \right\rangle }

\newcommand{\Um}{U_{\rm m}}
\newcommand{\Rem}{Re_{\rm m}}